\documentclass[nofootinbib,a4paper,aps,pra,showpacs,preprintnumbers,twocolumn]{revtex4}
\usepackage{cmap}
\usepackage{graphics,graphicx,epsfig}
\usepackage{epstopdf}
\usepackage[centertags]{amsmath}
\usepackage{amsfonts}
\usepackage{amssymb}
\usepackage[english]{babel}
\usepackage{amsthm,color}
\usepackage{euscript}
\DeclareGraphicsExtensions{.pdf,.png,.jpg,.eps}
\usepackage[T1]{fontenc}
\usepackage[utf8]{inputenc}
\usepackage{xcolor}
\usepackage{hyperref}
\definecolor{linkcolor}{HTML}{000000}
\definecolor{urlcolor}{HTML}{000000}
\hypersetup{pdfstartview=FitH, linkcolor=linkcolor,urlcolor=urlcolor, colorlinks=true}

\begin{document}
  
\title{Teleportation protocols with non-Gaussian operations: conditional photon subtraction versus cubic phase gate}
\author{E.R. Zinatullin}
\author{S. B. Korolev}
\author{T. Yu. Golubeva}
\affiliation{St. Petersburg State University, Universitetskaya nab. 7/9, St. Petersburg, 199034, Russia}
\begin{abstract}
In our work, we compare three teleportation protocols: the original protocol, the photon subtraction protocol, and the protocol with a cubic phase gate. We evaluate the fidelity of each protocol using the example of teleportation of the squeezed state and the Schrödinger’s cat state. We show that, under equal conditions, the teleportation scheme with a cubic phase gate achieves significantly higher fidelity than the other protocols considered.
\end{abstract}
\maketitle
  

\section{Introduction}

Quantum teleportation is one of the basic protocols of quantum information processing \cite{Bennett,Vaidman,Bouwmeester,Braunstein,Furusawa}. It is this protocol that underlies one of the promising models of universal quantum computation - one-way quantum computation model \cite{Menicucci,Raussendorf,Nielsen}.  In our work, we will discuss the continuous-variable quantum teleportation protocol \cite{Lloyd,Braunstein2}. Unlike discrete quantum systems, the use of continuous-variable ones allows one to build deterministic schemes. However, working with continuous-variable quantum systems also has a significant drawback: the presence of unavoidable errors associated with the finite squeezing degree of states, which are used as a resource for teleportation. It is these errors that are the main limiting factor of the regime in question.

Continuous-variable one-way quantum computation has inherited this disadvantage. The squeezing degree, which is experimentally achievable at the moment, turns out to be insufficient for performing universal fault-tolerant quantum computations.  The maximum experimentally achievable squeezing degree is -15 dB \cite{Vahlbruch}, whereas for such computations (without using a post-selection procedure) a squeezing of -20.5 dB \cite{Menicucci1} is required. There are various approaches to circumvent the limitations of insufficient squeezing. These approaches include the use of post-selection \cite{Fukui} and surface codes \cite{Fukui,Noh,Fukui1,Noh1,Larsen1,Bourassa,Tzitrin}. For example, in \cite{Fukui} the authors proposed a computation scheme that allows reducing the squeezing requirements to -10.8 dB. Thus, the main efforts are usually aimed at error correction. However, the resource state requirements can be lowered by using computational schemes that are less sensitive to the initial error. The first step for building such schemes is to modify the basic one-way quantum computation protocol - the teleportation protocol.

One of the recipes to improve teleportation accuracy is to use a non-Gaussian state obtained by the conditional subtraction or addition of photons (PS) procedure as a teleportation resource. This method has been proposed in \cite{Opatrny}. However, such a modified protocol loses determinacy because of the probabilistic nature of the non-Gaussian operations used. Another teleportation scheme, described by us in \cite{Zinatullin1}, uses as a resource the non-Gaussian state obtained with a cubic phase gate (CPG) \cite{GKP}. In contrast to the scheme with PS, it works in a deterministic way.

 The first idea of generating cubic phase states was proposed by Gottesman, Kitaev, and Preskill back in 2001 \cite{GKP,Ghose,Gu}. It turned out that this idea is difficult to implement in practice since it requires performing the quadrature displacement operation by a value far from what is achievable in an experiment. Because of this, the CPG has long remained just an abstract mathematical transformation. However, the situation has changed in recent years.  There are more and more works devoted to new methods for the cubic phase states generation \cite{Yukawa,YZhang,Asavanant} and the implementation of CPG \cite{Hillmann,Marshall,Miyata,Yanagimoto,Konno}. Particularly significant advances have been made in the microwave frequency range - it was in this range that the cubic phase state was generated for the first time \cite{Kudra}.  As a result, the CPG gradually turns from a purely theoretical transformation into a real-life device. 

Thus, we can talk about the advantage of the scheme with CPG over the scheme with PS in terms of preserving the transformations’ determinism. However, the question arises, which of these schemes allows to perform teleportation better and gives a greater gain over the scheme using Gaussian resource. It is natural to consider teleportation fidelity as a measure of this comparison.

In our work, we will compare the original teleportation protocol \cite{Braunstein}, the teleportation protocol with PS \cite{Opatrny}, and the teleportation protocol with CPG \cite{Zinatullin1}. For this purpose, we briefly describe each of the protocols in Section \ref{SecProt}. Then in Section \ref{SecF} we will evaluate the fidelity for each of the protocols when teleporting Gaussian and non-Gaussian states. As the Gaussian state, we will consider the squeezed state, and the non-Gaussian state will be the Schrödinger cat state. Such comparison will allow to estimate, which of non-Gaussian procedures has more perspectives for introduction into one-way quantum computation schemes.


\section{Teleportation protocols} \label{SecProt}

Before we compare protocols, let us recall how each of them is constructed. In this section, we recall how the original continuous-variable teleportation protocol works. Then we will consider its modification with the PS procedure. Finally, we briefly describe the protocol of teleportation with CPG.

\subsection{Original teleportation protocol}

\begin{figure}[t]
\begin{center}
\includegraphics[width=75mm]{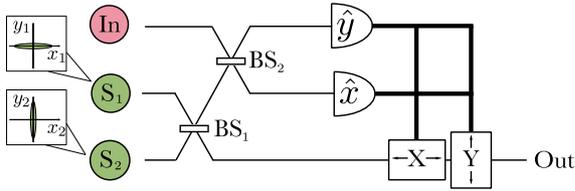}
\caption{The teleportation scheme of the input state In. On the scheme: $\text{S}_1$ and $\text{S}_2$ are resource oscillators squeezed in orthogonal quadratures; $\text{BS}_1$ and $\text{BS}_2$ are symmetric beam splitters; $\hat y$ and $\hat x$ are homodyne detectors, measuring the corresponding field quadratures in channels, X and Y denote devices that displace the corresponding field quadratures in the channel depending on the detection results.}
\label{ShemO}
\end{center}
\end{figure}

We begin our discussion by recalling how the original continuous-variable teleportation protocol is constructed. The two oscillators denoted as $S_1$ and $S_2$ in Fig. \ref{ShemO}, are squeezed in orthogonal directions. We will describe them by the following quadrature components:
\begin{align}
& \hat{x}_1=e^r \hat x_{0,1},\qquad\hat{y}_1=e^{-r} \hat y_{0,1}\label{s1},\\ 
& \hat{x}_2=e^{-r} \hat x_{0,2},\qquad\hat{y}_2=e^{r} \hat y_{0,2},\label{s2}
\end{align} 
where $\hat x_{0,j}$ and $\hat y_{0,j}$ are quadratures of the $j$-th oscillator in the vacuum state. The oscillators used are believed to be equally squeezed and the parameter $r$ specifies their squeezing degree. 

The squeezed fields are mixed on the symmetric beam splitter, which leads to the creation of the entangled state: 
 \begin{align}
&\hat{a}_1^\prime = \frac{1}{\sqrt{2}}\left(\left(\hat{x}_1+\hat{x}_2\right)+i\left(\hat{y}_1+\hat{y}_2\right)\right),\\
& \hat{a}_2^\prime = \frac{1}{\sqrt{2}}\left(\left(\hat{x}_1-\hat{x}_2\right)+i\left(\hat{y}_1-\hat{y}_2\right)\right).
 \end{align}
The resulting entangled state acts as a quantum resource for further teleportation.

Then the input (teleportable) state is mixed with the field in the first channel using a symmetrical beam splitter. As a result, the field operators will take the form:
\begin{align}
\hat{a}_{in}^\prime = & \frac{1}{\sqrt{2}}\Bigg(\left(\hat{x}_{in}+  \frac{1}{\sqrt{2}}\left(\hat{x}_1+ \hat{x}_2 \right)\right) \nonumber\\
&+i\left(\hat{y}_{in}+  \frac{1}{\sqrt{2}}\left(\hat{y}_1+\hat{y}_2\right)\right)\Bigg),
\end{align}
\begin{align}
\hat{a}_{1}^{\prime\prime} = & \frac{1}{\sqrt{2}}\Bigg(\left(\hat{x}_{in}-  \frac{1}{\sqrt{2}}\left(\hat{x}_1+ \hat{x}_2 \right)\right) \nonumber\\
&+i\left(\hat{y}_{in}-  \frac{1}{\sqrt{2}}\left(\hat{y}_1+\hat{y}_2\right)\right)\Bigg).
\end{align}

Next, using the homodyne detection procedure, we measure the $y$-quadrature of the field in the input channel and the $x$-quadrature in the first channel. Such a measurement, because of the entanglement of the resource state, will lead to a change in the quadratures of the field in the second channel:
\begin{align}
& \hat x_2^\prime = \hat x_{in} - \sqrt{2}\hat x_{s,2}- \sqrt{2}X_{1}, \label{x2'}\\
& \hat y_2^\prime = \hat y_{in} + \sqrt{2}\hat y_{s,1}- \sqrt{2}Y_{in}. \label{y2'}
\end{align}
Here $X_{1}$ and $Y_{in}$ are measured quadrature values.

Finally, the last step in the teleportation protocol is to displace quadratures in the second channel. One should choose the displacement value as so to compensate for the c-number terms in the Eqs. (1)-(2). As a result, the state at the output of the scheme takes the form:
\begin{align}
& \hat x_{out} = \hat x_{in} - \sqrt{2}\hat x_{s,2} = \hat x_{in} - \sqrt{2}e^{-r} \hat x_{0,2},\label{10}\\
& \hat y_{out} = \hat y_{in} + \sqrt{2}\hat y_{s,1} = 
 \hat y_{in} + \sqrt{2}e^{-r} \hat y_{0,1},\label{11}
\end{align}
where the second equalities consider the squeezing degree of the resource oscillators (\ref{s1})-(\ref{s2}). Thus, the output quadratures are equal to the input ones with the addition of errors from the non-ideally squeezed quadratures of the resource oscillators.

The above reasoning is convenient for a clear demonstration of the protocol operation. However, it does not allow us to evaluate the quality of teleportation for specific input states. Therefore, we will once again consider the original teleportation protocol, but in Schrödinger’s representation. In the following, we will repeat the reasoning outlined by the authors in the article \cite{Opatrny}.

Let the input state be described by the wave function in the coordinate representation $\psi_{in}(x_{in})$, and the entangled resource oscillators (i.e. after the first beam splitter) be described by the wave function $\psi_{1,2}(x_1,x_2)$. The second beam splitter acts on the oscillators' quadrature as
\begin{align}
\hat x_{in} \to \frac{\hat x_{in}+\hat x_1}{\sqrt{2}}, \quad
\hat x_1 \to \frac{\hat x_{in}-\hat x_1}{\sqrt{2}}.
\end{align}
The wave function of the system is transformed as follows
\begin{align}
\psi_{in,1,2}(x_{in},x_1,x_2)=&\psi_{in}\left(\frac{x_{in}+x_1}{\sqrt{2}}\right)\psi_{1,2} \nonumber\\
&\times\left(\frac{x_{in}-x_1}{\sqrt{2}},x_2\right).
\end{align}
Then the quadratures $x_1$ and $y_{in}$ are measured. The unnormalized wave function of the second oscillator after such a measurement takes the form
\begin{align}
\psi_2(x_2;Y_{in},X_1)=&\frac{1}{\sqrt{\pi}}\int dx_{in} \, e^{-2iY_{in}x_{in}}\psi_{in}\left(\frac{x_{in}+X_1}{\sqrt{2}}\right) \nonumber\\
&\times\psi_{1,2}\left(\frac{x_{in}-X_1}{\sqrt{2}},x_2\right).
\end{align}
The probability density that the measurement of the quadratures $\hat x_1$ and $\hat y_{in}$ will yield values $X_{1}$ and $Y_{in}$ is determined by the following expression:
\begin{align}
P(Y_{in},X_1)=\int  dx_2 \, |\psi_2(x_2;Y_{in},X_1)|^2.
\end{align}
To complete the teleportation procedure, it remains to displace the quadratures in the second channel by the measured quadrature values. Thus, the wave state function at the output of the scheme is given by the expression
\begin{align}
\psi_{out}(x;Y_{in},X_1)=\frac{1}{\sqrt{\pi P(Y_{in},X_1)}}\int dx_{in} \, e^{2iY_{in}(\sqrt{2}x-x_{in})} \nonumber\\
\times\psi_{in}\left(\frac{x_{in}+X_1}{\sqrt{2}}\right)\psi_{1,2}\left(\frac{x_{in}-X_1}{\sqrt{2}},x+\sqrt{2}X_1\right). \label{psiOut0}
\end{align}

Next, we will need to modify this protocol by adding a conditional photon subtraction procedure. With this in mind, it will be convenient to proceed to the decomposition of the wave functions by the set of Fock states:
\begin{align}
\varphi_k(x)=\frac{1}{\sqrt{2^{k-1}k!\sqrt{\pi}}} e^{-x^2} H_k(\sqrt{2} x), \quad k=0,1,2...\, .\label{phi_fock}
\end{align}
In Eq. (\ref{phi_fock}) $H_k(x)$ are Hermite polynomials. Then, the wave function of the entangled resource oscillators can be represented as
\begin{align}
\psi_{1,2}(x_1,x_2)=\sum_k a_{k} \varphi_k(x_1) \varphi_k(x_2), \label{psi12_O}
\end{align}
where the coefficients have the form
\begin{align}
a_{k}=\sqrt{1-q^2} q^k.
\end{align}
In this decomposition, the parameter $q$ ($0<q<1$) is responsible for an entanglement strength and is related to the oscillators’ squeezing as:
\begin{align}
q=\tanh r
\end{align}
We also represent the wave function of the teleported input oscillator as the sum
\begin{align}
\psi_{in}(x_{in})=\sum_m a_{m}^{in} \varphi_m(x_{in}). \label{psiIN}
\end{align}

If we substitute the decompositions (\ref{psi12_O}) and (\ref{psiIN}) into the Eq. (\ref{psiOut0}), we can write the expression for the wave function of the output state as
\begin{align}
\psi_{out}(x;Y_{in},X_1)&=\frac{1}{\sqrt{\pi P(Y_{in},X_1)}} e^{2\sqrt{2}i Y_{in} x}\sum_{k,m} a_k a_{m}^{in} \nonumber\\
&\times\varphi_k(x+\sqrt{2}X_1) D_{k,m}(Y_{in},X_1). \label{Out_t}
\end{align}
Here we introduced the c-numeric function $D_{k,m}(Y_{in},X_1)$ defined by expressions:
\begin{align}
D_{k,m}(Y_{in},&X_1) =2\sqrt{2^{m-k}\frac{k!}{m!}} e^{-Y_{in}^2-X_1^2} \nonumber\\
&\times(X_1 - i Y_{in})^{m-k} L_k^{m-k}(2 Y_{in}^2+2 X_1^2) \label{func_D}
\end{align}
for $m \geqslant k$, and 
\begin{align}
D_{m,k}(Y_{in},X_1) =(-1)^{m-k} D_{k,m}^*(Y_{in},X_1)
\end{align}
for $m < k$. In Eq. (\ref{func_D}) $L_k^m(x)$ are generalized Laguerre polynomials. The resulting form of the transformation will be convenient for modifying the scheme by PS.


\subsection{Teleportation protocol with PS}

\begin{figure}[t]
\begin{center}
\includegraphics[width=85mm]{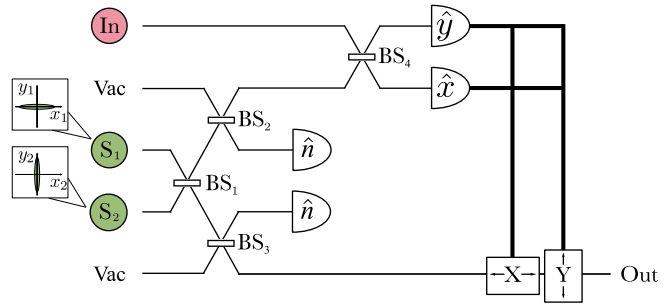}
\caption{The teleportation scheme with the PS. On the scheme: open channels are designated as Vac; $\text{BS}_1$ and $\text{BS}_4$ are symmetric beam splitters; $\text{BS}_2$ and $\text{BS}_3$ are low-reflectance beam splitters; $\hat y$ and $\hat x$ are homodyne detectors, and $\hat {n}$ are photon-number detectors}
\label{ShemPS}
\end{center}
\end{figure}

In \cite{Opatrny}, the authors proposed a modified teleportation protocol (see Fig. \ref{ShemPS}). In this protocol, the entangled non-Gaussian state is used as a resource. This state is obtained from the two-mode squeezed vacuum by subtracting photons from each mode. The conditional PS procedure is performed as follows: a beam splitter with small amplitude reflection coefficient $r$ (and amplitude transmittance $t$) is placed in the channel, then the number of photons is measured in the reflected beam. Such an operation on the $j$-th oscillator transforms the Fock state $|k_j\rangle$ as follows:
\begin{align}
|k_j\rangle \to (-1)^{n_j}\sqrt{\frac{(k_j+n_j)!}{k_j! n_j!}}|r|^{n_j} |t|^{k_j-n_j} |k_j-n_j\rangle,
\end{align}
where $n_j$ is the number of measured photons in the reflected beam (see \cite{Dakna1,Dakna2}).

A successful implementation of the protocol will be considered the case when one photon has been detected in each channel. Then, the non-Gaussian entangled resource state can be represented as the decomposition:
\begin{align}
\psi_{1,2}^{ps}(x_1,x_2)=\sum_k a_{k}^{ps} \varphi_k(x_1) \varphi_k(x_2),
\end{align}
where the coefficients are given by the following relation:
\begin{align}
a_{k}^{ps}=\sqrt{1-q^2}\frac{(k+1)!}{k!}|r|^2 |t|^{2k} q^{k+1}.
\end{align}
Since the rest of the scheme is the same as the original one, the Eq. (\ref{Out_t}) describes the output state up to changing $a_{k} \to a_{k}^{ps}$.


\subsection{Teleportation protocol with CPG}

\begin{figure}[h]
\begin{center}
\includegraphics[width=85mm]{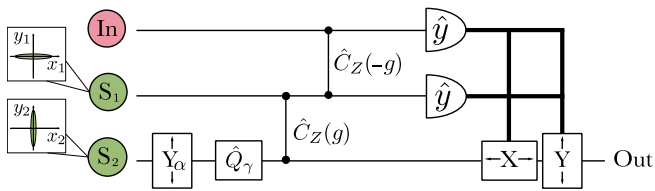}
\caption{The teleportation scheme with the CPG.  On the scheme: $\text{Y}_\alpha$ denotes the displacement of the $y$-quadrature by a fixed value $\alpha$; $\hat Q_\gamma$ is a cubic phase gate with a nonlinearity coefficient $\gamma$; $\hat{C}_{Z}(g) $ is a CZ transformation with weight coefficients $g$.}
\label{ShemQ}
\end{center}
\end{figure}

Let us now consider the teleportation protocol we proposed in \cite{Zinatullin1}. This protocol uses the entangled state modified by a cubic phase gate \cite{GKP} as a resource. Fig.\ref{ShemQ} shows the scheme of this protocol. As in the previous sections, we will describe the action of the protocol using wave functions in coordinate representation. But now we no longer need to proceed to the decomposition of the wave functions by the set of Fock states.

Each of the squeezed oscillators at the input of the scheme is described by a wave function:
\begin{align}
\psi_s(x;r_j)=\sqrt[4]{\frac{2 e^{2r_j}}{\pi}}\exp\left(-e^{2r_j} x^2\right),
\end{align}
where $r_j$ is the squeezing coefficient of the corresponding oscillator. In our protocol $r_2=-r_1=r$. Next, a non-Gaussian state is prepared by sequentially applying to the second squeezed oscillator the displacement procedure by the value of $\alpha>0$:
\begin{align}
\hat Y_{\alpha,2}=e^{2i \alpha \hat x_2}
\end{align}
and CFG
\begin{align}
\hat Q_{\gamma,2}=e^{-2i\gamma \hat y_2^3},
\end{align}
where $\gamma$ is the nonlinearity coefficient. After these operators act, the wave function of the second oscillator takes the form: 
\begin{align}
\psi_{2}(x_2)=\frac{1}{\sqrt{\pi}} \int dy_2 \, e^{2i y_2 (x_2-\gamma y_2^2)} \psi_s(y_2-\alpha;-r),
\end{align}

In contrast to previous teleportation protocols, for entanglement we use the CZ transformation (see, for example \cite{Larsen,Su2018,Alexander}), which acts on the $j$-th and $k$-th oscillators as
\begin{align}
\hat C_{z,jk}(g)=e^{2i g \hat x_j \hat x_k}.
\end{align}
Here $g$ is the weight coefficient, which can be any positive or negative value. It was shown in \cite{Zinatullin2} that using such an entanglement transformation instead of the beam splitter transformation reduces the teleportation error of one quadrature by a factor of g. We apply two CZ transformations sequentially. The first CZ transformation entangles the resource oscillators, and the second one entangles the input state with the state in the first channel. After that, the wave function describing the whole system has the form 
\begin{align}
\psi_{in,1,2}(x_{in},x_1,x_2)=&e^{2ig x_1 (x_2-x_{in})} \psi_s(x_1;-r) \nonumber\\
&\times\psi_{2}(x_2) \psi_{in}(x_{in}).
\end{align}

Next, we measure the $y$-quadratures of the input and the first oscillators. We will consider the measured values to be $Y_{in}$ and $Y_{1}$, respectively. Homodyne measurement is a projection operation on the $y$-quadratures’ eigenstates, corresponding to the measured values of the photocurrents. Then the unnormalized wave function of the second oscillator after the measurement takes the form of 
\begin{align}
\psi_2'(x_2;Y_{in},Y_1)=&\frac{1}{\sqrt{\pi}}\int  dx_{in} \, e^{-2ix_{in}Y_{in}} \psi_{2}(x_2) \psi_{in}(x_{in}) \nonumber\\
&\times \psi_s\big(g(x_2-x_{in}-Y_1/g);r\big).
\end{align}
The resulting wave function must be normalized by the square root of the probability density that the measurement of the quadratures $\hat y_1$ and $\hat y_{in}$ will yield values $Y_{1}$ and $Y_{in}$:
\begin{align}
P(Y_1,Y_{in})=\int  dx_2 \, |\psi_2'(x_2;Y_{in},Y_1)|^2.
\end{align}

To complete the teleportation procedure, the $x$-quadrature of the second oscillator should be displaced by $-Y_1/g$, and the $y$-quadrature by $Y_{in}-\sqrt{Y_1/(3 \gamma g)}$. Thus, the wave function of the teleported state has the following form:
\begin{align}
\psi_{out}&(x;Y_{in},Y_1)=\frac{1}{\sqrt{\pi P(Y_1,Y_{in})}}\int  dx_{in} \, e^{-2ix_{in} Y_{in}} \nonumber\\
&\times \exp\left(2i \left(Y_{in}-\sqrt{\frac{Y_1}{3\gamma g}}\right) \left(x+\frac{Y_1}{g}\right)\right) \nonumber\\
&\times \psi_s\big(g(x-x_{in});r\big) \psi_{2}\left(x+\frac{Y_1}{g}\right) \psi_{in}(x_{in}).
\end{align}

\begin{figure*}[t]
\begin{center}
\includegraphics[width=170mm]{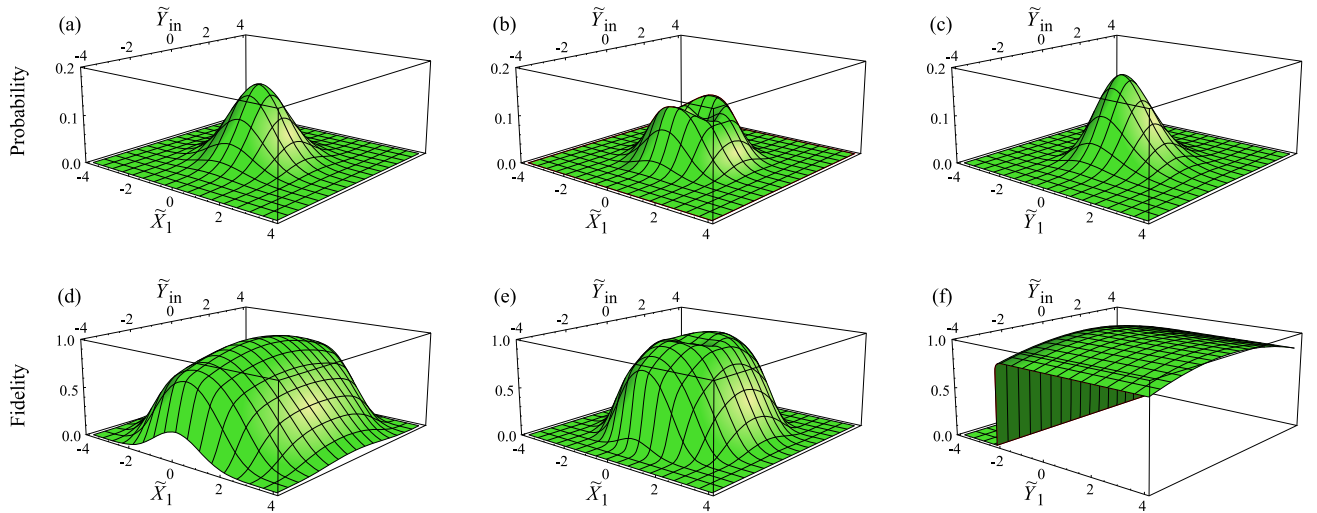}
\caption{The probability densities of measuring the values $\tilde X_1$ (or $\tilde Y_1$) and $\tilde Y_{in}$ (a,b,c) as well as teleportation fidelity (d,e,f) for squeezed state: (a,d) original teleportation protocol; (b,e) teleportation with PS (c,f) teleportation with CPG.}
\label{PFsq}
\end{center}
\end{figure*}

The paper \cite{Zinatullin1} presents a detailed analysis of this protocol. 


\section{Comparison of different teleportation protocols} \label{SecF}

Let us evaluate the performance of each protocol for specific input states. To do this, we calculate the teleportation fidelity, which is defined as
\begin{align}
F(Y_{in},E_1)=\left| \int dx \, \psi_{out}^*(x;Y_{in},E_1) \psi_{in}(x)\right|^2.
\end{align}
Here, for the original teleportation protocol and the teleportation with PS, the quadrature measured in the first channel is $X_1$ (i.e., $E_1=X_1$), and for the teleportation with QPG $E_1=Y_1$.

The protocols we consider have significantly different working areas (i.e. different ranges of measured quadrature values). To compare fidelity graphs for different protocols, it is necessary to combine these areas. To do this, we center and normalize the axes of the measured quadratures. We will normalize them by the variances values of the corresponding quadratures.
\begin{align}
\tilde E_j=\frac{E_j-\langle E_j\rangle}{\langle (E_j-\langle E_j\rangle)^2\rangle}, \quad E_j=Y_{in},X_1,Y_1.
\end{align}
Such a scaling seems natural, since the first and second moments of the probability densities $P(\tilde Y_{in}, \tilde E_1)$ are equal, and their differences are due only to non-Gaussian features of the protocols themselves and the teleported states.

Let us briefly discuss the parameters we took for the calculations. We will take the squeezing of resource oscillators for all protocols the same and equal to $-10$ dB. Such squeezing is achievable in the real experiment \cite{Vahlbruch}. For a teleportation protocol with PS, a balance between two factors should be maintained when selecting the reflection coefficient of the beam splitters used for photon subtraction. On the one hand, the reflection coefficient should be small so that cases of subtraction of two or more photons can be neglected (since existing photon detectors do not allow us to measure the exact number of photons that have arrived). On the other hand, it cannot be very small, so that the probability of successful protocol implementation does not tend to zero. Following the authors of \cite{Opatrny}, we take $r=0.05$. In this case, the probability of successful PS in both channels is approximately 4\%. Now let us turn to the parameters used in the teleportation protocol with CPG. Usually, an auxiliary squeezed oscillators (see, for example \cite{Zinatullin2}) are required to implement the CZ transformation. We assume that the squeezing of these oscillators is the same as that of the resource ones (i.e., $-10$ dB). Then the weight coefficient of the CZ transformation can be taken approximately $g=3$ \cite{Zinatullin3}. We take the relatively small nonlinearity of the CPG $\gamma=0.1$, and the displacement $\alpha=7$. Small values of nonlinearity are estimated as conditionally achievable \cite{Yukawa, Kudra}, and displacement by a small value does not represent experimental difficulties \cite{Zinatullin1}.

\begin{figure*}[t]
\begin{center}
\includegraphics[width=170mm]{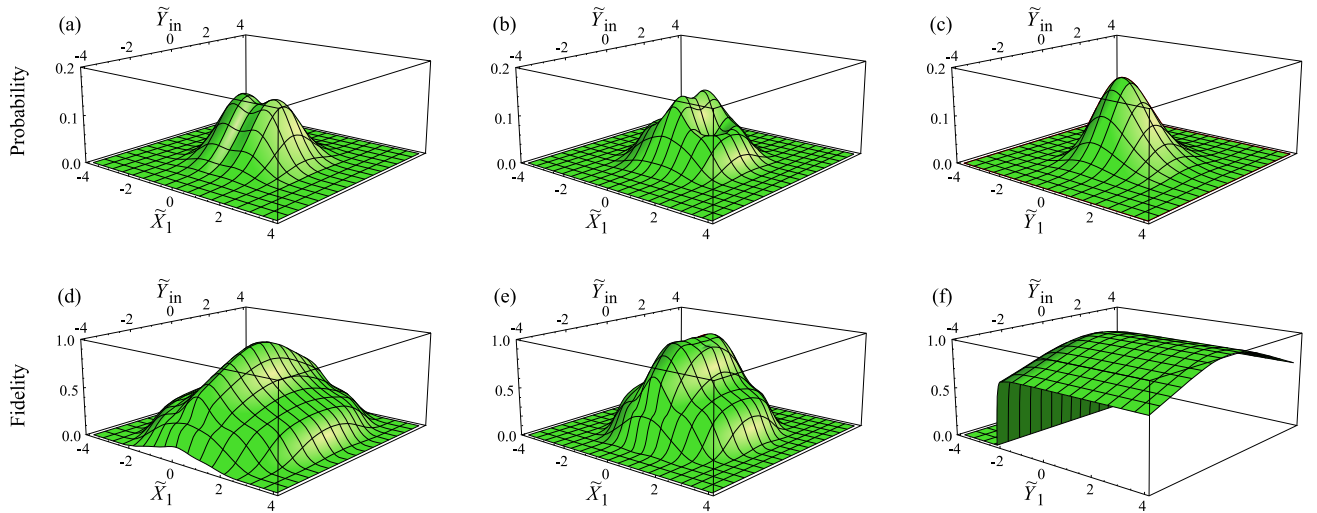}
\caption{The probability densities of measuring the values $\tilde X_1$ (or $\tilde Y_1$) and $\tilde Y_{in}$ (a,b,c) as well as teleportation fidelity (d,e,f) for Schrödinger cat state: (a,d) original teleportation protocol; (b,e) teleportation with PS (c,f); teleportation with CPG.}
\label{PFcat}
\end{center}
\end{figure*}

\subsection{Squeezed state teleportation}

It is of interest to evaluate for which type of states the use of a non-Gaussian resource gives a greater gain and which of the protocols provides it. To evaluate the performance of the protocols, we considered the cases of Gaussian and non-Gaussian input states. As a Gaussian one, we took a $x$-quadrature squeezed state with $-5$ dB squeezing. 

Fig. \ref{PFsq} shows the probability density of measuring the values $\tilde X_1$ (or $\tilde Y_1$) and $\tilde Y_{in}$ as well as teleportation fidelity of Gaussian state for all analyzed protocols. Let us first compare the original protocol and the protocol with PS. For the original protocol, the probability density $P(\tilde Y_{in},\tilde X_1)$ has the form of a Gaussian distribution. For the teleportation protocol with PS, the probability density $P(\tilde Y_{in},\tilde X_1)$ has a dip in the center of the working area. Comparing the fidelities for these protocols, we see that the fidelity for the protocol with PS is slightly higher in the area of the most probable measured values, but less at the edges of this area. Therefore, it is difficult to say unequivocally which of the protocols turned out to be more effective. At the same time, the probability density $P(\tilde Y_{in},\tilde Y_1)$ for the protocol with CPG is asymmetric about the $\tilde Y_{in}$-axis. It has a tail going into the area with high fidelity values. Furthermore, the fidelity of the teleportation protocol with CPG exceeds the fidelity of the other protocols in almost the entire working area.

\subsection{Schrödinger's cat state teleportation}

As a non-Gaussian input state, we took an odd Schrödinger cat state, which is a superposition of two coherent states:
\begin{align}
\psi_{cat}(x)=\frac{1}{N_{cat}}\left(e^{2ibx}-e^{-2ibx}\right)e^{-x^2},
\end{align}
where $N_{cat}$ is the normalization factor. For the calculations, we take $b=1.5$. Using the same state, the authors of the paper \cite{Opatrny} tested the operation of the protocol with PS.

Fig. \ref{PFcat} shows the probability density of measuring the values $X_1$ (or $Y_1$) and $Y_{in}$ as well as the teleportation fidelity for the three protocols considered. When comparing the original teleportation protocol and the protocol with PS, we see that the probability density $P(\tilde Y_{in},\tilde X_1)$ for the protocol with PS increases in the central part of the working area. Also, for the chosen non-Gaussian input state, the fidelity of the teleportation with PS increased significantly in the central part of the work area. Thus, when teleporting the Schrödinger cat state, the protocol with PS is more efficient than the original one. However, as well for the squeezed input state, the fidelity for the protocol with CPG is higher than for the others in almost the entire working area. For protocol with CPG, the probability density $P(\tilde Y_{in},\tilde Y_1)$ for Schrödinger’s cat teleportation and squeezed state teleportation has the same form.

\vspace{0.5 cm}

\begin{figure}[h]
\begin{center}
\includegraphics[width=80mm]{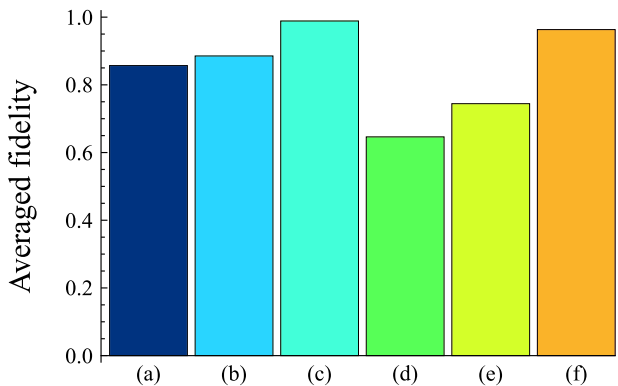}
\caption{Averaged teleportation fidelity: (a) squeezed state teleportation by the original protocol; (b) squeezed state teleportation by the protocol with PS; (c) squeezed state teleportation by the protocol with CPG; (d) Schrödinger's cat state teleportation by the original protocol; (e) Schrödinger's cat state teleportation by the protocol with PS; (f) Schrödinger's cat state teleportation by the protocol with CPG.}
\label{Fint}
\end{center}
\end{figure}

To objectively compare the quality of all three protocols, for each of the cases considered above, we calculate the averaged fidelity of teleportation:
\begin{align}
\langle F \rangle=\iint dY_{in} dE_1 \, F(Y_{in},E_1).
\end{align}
The calculation results are shown in Fig. \ref{Fint}. We see that the averaged fidelity of squeezed state teleportation for the protocol with PS is slightly higher than for the original protocol. The gain for the protocol with PS becomes more noticeable when teleporting the more complex Schrödinger’s cat state. However, the averaged fidelity for the protocol with CPG significantly exceeds the averaged fidelity for the other protocols for both the Gaussian input state and the non-Gaussian one.

Thus, under the same conditions, the teleportation protocol with CPG is more efficient than the other protocols considered. In addition, it works deterministic, while for the protocol with PS the probability of successful implementation is only 4\%.


\section{Conclusion}

In our work, we have compared the performance of three teleportation protocols: the original protocol, the protocol with PS, and the protocol with CPG. On the example of squeezed state and Schrödinger’s cat state, we have shown that the protocol with CPG allows to reach higher fidelity values. It should be noted that teleportation fidelity in the scheme with CPG is almost independent of the type of teleported state. This distinguishes it noticeably from other protocols. In addition, the protocol with CPG works in a deterministic way, unlike the protocol with PS.

It is worth noting that the implementation of the teleportation protocol with the CPG is technically more difficult than implementing the other protocols considered. The key element of this scheme is the CPG, the practical implementation of which is still a challenge for experimenters. In addition, the protocol assumes confusion of the fields using CZ transformations. Such entanglement is more difficult to implement in practice than entanglement on a beam splitter (see, for example, \cite{Zinatullin2}). Nevertheless, Gaussian and non-Gaussian resources are currently being actively developed. In addition, it has made significant progress in the generation of cubic phase states \cite{Kudra}. Therefore, with the advent of better ways to implement the CPG and CZ transformation, protocol with the cubic phase gate can give a significant advantage over other teleportation protocols.

We can conclude that the CPG has more prospects for further implementation in one-way quantum computation schemes.

\vspace{0.5 cm}

This research was supported by the Theoretical Physics and Mathematics Advancement Foundation "BASIS" (Grants No. 21-1-4-39-1 and No. 22-1-5-90-1 ).




\begin{thebibliography}{100}
%
\bibitem{Bennett} C. H. Bennett, G. Brassard, C. Crépeau, R. Jozsa, A. Peres and W. K. Wootters, Phys. Rev. Lett. {\bf 70}, 1895 (1993).
%
\bibitem{Vaidman} L. Vaidman, Phys. Rev. A {\bf 49}, 1473 (1994).
%
\bibitem{Bouwmeester} D. Bouwmeester, J.-W. Pan, K. Mattle, M. Eibl, H. Weinfurter and A. Zeilinger, Nature {\bf 390}, 575 (1997).
%
\bibitem{Braunstein} S. L. Braunstein and H. J. Kimble, Phys. Rev. Lett. {\bf 80}, 869 (1998).
%
\bibitem{Furusawa} A. Furusawa, J. L. Sørensen, S. L. Braunstein, C. A. Fuchs, H. J. Kimble and E. S. Polzik, Science {\bf 282}, 706 (1998).
%
\bibitem{Menicucci} N. C. Menicucci, P. van Loock, M. Gu, C. Weedbrook, T. C. Ralph and M. A. Nielsen, Phys. Rev. Lett. {\bf 97}, 110501 (2006).
%
\bibitem{Raussendorf} R. Raussendorf and H. J. Briegel, Phys. Rev. Lett. {\bf 86}, 5188 (2001).
%
\bibitem{Nielsen} M. A. Nielsen, Reports on Mathematical Physics {\bf 57}, 147 (2006).
%
\bibitem{Lloyd} S. Lloyd and S. L. Braunstein, Phys. Rev. Lett. {\bf 82}, 1784 (1999).
%
\bibitem{Braunstein2} S. L. Braunstein and P. van Loock, Rev. Mod. Phys. {\bf 77}, 513 (2005).
%
\bibitem{Vahlbruch} H. Vahlbruch, M. Mehmet, K. Danzmann, and R. Schnabel, Phys. Rev. Lett. {\bf 117}, 110801 (2016).
%
\bibitem{Menicucci1} N. C. Menicucci, Phys. Rev. Lett. {\bf 112}, 120504 (2014).
%
\bibitem{Fukui} K. Fukui, A. Tomita, A. Okamoto and K. Fujii, Phys. Rev. X {\bf 8}, 021054 (2018).
%
\bibitem{Noh} K. Noh and C. Chamberland, Phys. Rev. A {\bf 101}, 012316 (2020).
%
\bibitem{Fukui1} K. Fukui, arXiv preprint arXiv:1906.09767.
%
\bibitem{Noh1} K. Noh, C. Chamberland, and F. G.S.L. Brandão, PRX Quantum {\bf 3}, 010315 (2022).
%
\bibitem{Larsen1} M. V. Larsen, C. Chamberland, K. Noh, J. S. Neergaard-Nielsen, and U. L. Andersen, PRX Quantum {\bf 2}, 030325 (2021).
%
\bibitem{Bourassa} J. E. Bourassa, R. N. Alexander, M. Vasmer, A. Patil, I. Tzitrin, T. Matsuura, D. Su, B. Q. Baragiola, S. Guha, G. Dauphinais, K. K. Sabapathy, N. C. Menicucci and I. Dhand, Quantum {\bf 5}, 392 (2021).
%
\bibitem{Tzitrin} I. Tzitrin, T. Matsuura, R. N. Alexander, G. Dauphinais, J. E. Bourassa, K. K. Sabapathy, N. C. Menicucci and I. Dhand, PRX Quantum {\bf 2}, 040353 (2021).
%
\bibitem{Opatrny} T. Opatrný, G. Kurizki and D.-G. Welsch, Phys. Rev. A {\bf 61}, 032302 (2000).
%
\bibitem{Zinatullin1} E. R. Zinatullin, S. B. Korolev, and T. Yu. Golubeva, Phys. Rev. A {\bf 104}, 032420 (2021).
%
\bibitem{GKP} D. Gottesman, A. Kitaev, and J. Preskill, Phys. Rev. A {\bf 64}, 012310 (2001).
%
\bibitem{Ghose} S. Ghose and B. C. Sanders, Journal of Modern Optics, {\bf 54}, 855 (2007).
%
\bibitem{Gu} M. Gu, Christian Weedbrook, N. C. Menicucci, T. C. Ralph, and P. van Loock, Phys. Rev. A {\bf 79}, 062318 (2009).
%
\bibitem{Yukawa} M. Yukawa, K. Miyata, H. Yonezawa, P. Marek, R. Filip, and A. Furusawa, Phys. Rev. A {\bf 88}, 053816 (2013).
%
\bibitem{YZhang} Y. Zheng, O. Hahn, P. Stadler, P. Holmvall, F. Quijandría, A. Ferraro, and G. Ferrini, PRX Quantum {\bf 2}, 010327 (2021).
%
\bibitem{Asavanant} W. Asavanant, K. Takase, K. Fukui, M. Endo, J. Yoshikawa, and A. Furusawa Phys. Rev. A {\bf 103}, 043701 (2021).
%
\bibitem{Marshall} K. Marshall, R. Pooser, G. Siopsis, and C. Weedbrook, Phys. Rev. A {\bf 91}, 032321 (2015).
%
\bibitem{Miyata} K. Miyata, H. Ogawa, P. Marek, R. Filip, H. Yonezawa, J. Yoshikawa, and A. Furusawa, Phys. Rev. A {\bf 93}, 022301 (2016).
%
\bibitem{Yanagimoto} R. Yanagimoto, T. Onodera, E. Ng, L. G. Wright, P. L. McMahon, and H. Mabuchi, Phys. Rev. Lett. {\bf 124}, 240503 (2020).
%
\bibitem{Hillmann} T. Hillmann, F. Quijandría, G. Johansson, A. Ferraro, S. Gasparinetti, and G. Ferrini, Phys. Rev. Lett. {\bf 125}, 160501 (2020).
%
\bibitem{Konno} S. Konno, A. Sakaguchi, W. Asavanant, H. Ogawa, M. Kobayashi, P. Marek, R. Filip, J. Yoshikawa, and A. Furusawa, Phys. Rev. Applied {\bf 15}, 024024 (2021).
%
\bibitem{Kudra} M. Kudra, M. Kervinen, I. Strandberg, S. Ahmed, M. Scigliuzzo, A. Osman, D.P. Lozano, M.O. Tholén, R. Borgani, D.B. Haviland, G. Ferrini, J. Bylander, A.F. Kockum, F. Quijandría, P. Delsing, and S. Gasparinetti, PRX Quantum {\bf 3}, 030301 (2022).
%
\bibitem{Dakna1} M. Dakna, T. Anhut, T. Opatrný, L. Knöll, and D.-G. Welsch, Phys. Rev. A {\bf 55}, 3184 (1997).
%
\bibitem{Dakna2} M. Dakna, L. Knöll and D.-G. Welsch, Eur. Phys. J. D {\bf 3}, 295 (1998).
%
\bibitem{Larsen} M. V. Larsen, J. S. Neergaard-Nielsen, and U. L. Andersen, Phys. Rev. A {\bf 102}, 042608 (2020).
%
\bibitem{Su2018}D. Su, C. Weedbrook and K. Br\'adler, Phys. Rev. A \textbf{98}, 042304 (2018).
%
\bibitem{Alexander} R. N. Alexander, S. C. Armstrong, R. Ukai, and N. C. Menicucci, Phys. Rev. A \textbf{90}, 062324 (2014).
%
\bibitem{Zinatullin2} E. R. Zinatullin, S. B. Korolev, and T. Yu. Golubeva, Phys. Rev. A {\bf 103}, 062407 (2021).
%
\bibitem{Zinatullin3} E. R. Zinatullin, S. B. Korolev, A. D. Manukhova, and T. Yu. Golubeva, Phys. Rev. A {\bf 106}, 032414 (2022).

\end{thebibliography}
 \end{document}